\documentclass[12pt]{article}
\usepackage{epsfig}
\usepackage{listings}

\begin{document}
\thispagestyle{empty} 

\begin{flushright} 
\begin{tabular}{l} 
ANL-HEP-PR-01-025 \\
hep-th/0612158 \\ 
\end{tabular} 
\end{flushright}  

\vskip .3in 
\begin{center} 

{\Large\bf  Numerical Program for Computing $\Phi^3$ Amplitudes} 

\vskip .3in 

{\bf Gordon Chalmers} 
\\[5mm] 
{\em UCLA Physics Department \\ 
405 Hilgard Ave \\ 
Knudsen Hall\\ 
LA, CA  90095-1547} \\  

{e-mail: chalmers@physics.ucla.edu}  

\vskip .5in minus .2in 

{\bf Abstract}  
\end{center} 

A computing program in Matlab is given that computes amplitudes in 
scalar $\phi^3$ theory.  The program is partitioned into several parts 
and a simple guide is given for its use.  

\vfill\break 

The scattering amplitudes in $phi^3$ are computationally difficult at 
high loop orders due to the number of diagrams involved.  Their number 
grows as $(n-1)!$ at $n$-point and at tree level.  An automated program 
is given here that computes an arbitrary tree amplitude and then sews 
them together to obtain the quantum amplitudes.  The program is written 
in Matlab and is given in several pieces.  

The input parameters for the $n$-point amplitudes are given in the 
subprogram phi3compute.txt.  

The subprogram NMatrix.m must be loaded with the number of external 
lines ranging from npointlower to npoint.  These numbers span the 
numbers of external lines which are used in the trees that are sewn 
together into the loop amplitudes.  

The subprograms NodalComplex.m, treecall.m, innerproduct.m, and 
denomcall.m are saved as files in the directory to which the program 
can call.  

The program RainbowCompute.m is used to call the MonteCarlo simulation 
that computes the multi loop amplitude.  

The input variables are self-explanatory in the subpart Phi3Compute.txt.  
They are: kmomol, kmomor delimit the four-point kinematics to the four-point 
amplitude, gnumber is the power of the coupling constant, d is the 
dimension, kstep and sample are the step size to the partitions of momenta 
in the loop and sample is the number of samples of the internal momenta 
taken.  In the program RainbowCompute.m the variables are MonteRunSet denoting 
the number of samples, kmax is the discretized momentum, sampleX is the 
number of external momenta data points.  The output is delivered in the 
array RainbowCompute and there is a subroutine to plot the output.  

I typically call all of the programs and then paste in the subprogram 
RainbowCompute.m.  The time to obtain the output really varies depending 
on the parameters and the number of internal loops.  

Background algorithms to the computations are obtained in \cite{Chalmers1}, 
\cite{Chalmers2}.  The program is contained in the tex source following 
the bibliography.

\section*{Acknowledgements} 

The work of GC is supported in part by the US Department of Energy, 
Division of High Energy Physics, contract .

\clearpage
\appendix

\lstset{basicstyle=\scriptsize,breaklines=true}
\begin{lstlisting}
%%%%%%%%%%%%%%%%%%%%%%%%%%%%%%%%%%%%%%%%%%%%%%%%%%%
%%%%%%%%%%%%%%%%%%%%%%%%%%%%%%%%%%%%%%%%%%%%%%%%%%%
%%  
%%  Phi3Compute.txt
%% 

% momentum conservation at every node 
% check the redundance of the phi3 graphs


%%  this is the header to phi3 computation 

%%  be sure to preload the N matrix first with NMatrix(npoint) program


gnumber=6; 

ppoint=2; 

qpoint=2; 

d=4;

clear kmomol; 

clear kmomor; 

kstep=2; 

sample=100; 


for i=1:4

  kloffset(i)=.02323232323; 

  kroffset(i)=.01515151515;

end

kmomol(2,4,100)=-.004*kstep*sample/2; 
kmomor(2,4,100)=-.004*kstep*sample/2;


for samplex=1:sample 

for i=1:d

  for j=1:ppoint 

   kmomol(j,i,samplex)=kstep*rand+kloffset(j); 

  end 

end 


for i=1:d

   kmomol(1,i,samplex)=kloffset(i)+.004*kstep*samplex; 
   kmomol(2,i,samplex)=(1-kmomol(1,i,samplex)*kmomol(1,i,samplex))^(1/2); 


   kmomor(1,i,samplex)=kroffset(i)+.004*kstep*samplex; 
   kmomor(2,i,samplex)=(1-kmomor(1,i,samplex)*kmomor(1,i,samplex))^(1/2)-kmomol(1,i,samplex)-kmomor(1,i,samplex)-kmomor(2,i,samplex); 

end

end 


clear RainbowComputeOne;

RainbowComputeOne=RainbowCompute(sample,gnumber,ppoint,qpoint,kmomol,kmomor);

for j=1:100 

   testx(j)=j; 
   
end 

figure  

plot(testx,RainbowComputeOne)

%%%%%%%%%%%%%%%%%%%%%%%%%%%%%%%%%%%%%%%%%%%%%%%%%%%%%%%%%%%
%%
%%  RainbowCompute.m
%%
%%

function[RainbowCompute]=RainbowCompute(sample,gnumber,ppoint,qpoint,kmomol,kmomor)

% had to preload the Nmatrix containing the variable N 

% doesnt contain the asterisked data immediately below 

% function doesnt work so paste in the information 

% gnumber=10; 

% ppoint=2; 

% qpoint=2;

[lineconfig,NodeNumberPerm]=NodalComplex(gnumber,ppoint,qpoint); 

load Nmatrix

sampleX=100;

%% kmax is the max value of the momenta individual components in the integral

kmax=.01; 

% need an offset for the momentum to isolate uv from ir regime 

samplex=sample;

m=1;

d=4; 

% Two Nodes: InternalLines+ppoint-2+InternalLines+qpoint-2=gnumber  

InternalLinesMax=(gnumber+4-ppoint-qpoint)/2;  

InternalLinesMin=2;

NodeNumberMax=(gnumber-ppoint-qpoint+2)/2; 

%% the number of random samples in the integral 

RainbowCompute(sample)=0; 

RainbowComputetwo=0; 


%% number of external momenta data points 

sampleX=100; 

%% number of integral momenta data points 

MonteRunSet=100;

kmeasure=kmax/MonteRunSet; 

%%%%%%%%%%%%%%%%%  symmetrize in the momenta  %%%%%%%%%%%   

%% 

plr(ppoint+qpoint,d,100)=0;  

testvec(ppoint,d,100)=0; 


%% repeat many times to monte carlo simulate the integrals 

for samplex=1:sampleX
    
%% permutation set

 for permmomenta=1:3 
    
 % clear data 

  NodeNumber=1; 

  clear internallines; internallines(NodeNumberMax)=0;  

  clear InternalTest; InternalTest=0;     

  ltest=0; 


%% this evaluates all the graphs in the nodal complex 

for MonteRun=1:MonteRunSet;

for permno=1:NodeNumberPerm 

    
%%   j ranges from 1 to NodeNumber 
%%  

   clear internallinesval; 

   internallinesval(InternalLinesMax,d,NodeNumberMax,NodeNumberPerm)=0;

   RainbowIntegral(NodeNumberPerm)=1; 

   leftno=0; 

   rightno=0; 

   leftnodeno=0; 

   rightnodeno=0;

   maxrightno=1; 

   clear k; 
   
   
%% store InternalLinesVal(nodenumber,linenumber,dimension,permconfig) 
%% for all permutation sets contributing to the rainbow graphs 

   for j=1:NodeNumberPerm  

       clear momentumsum; 
       momentumsum(NodeNumberMax,d)=0;  
       
       clear NodeNumberMaxTwo=0;
       NodeNumberMaxTwo=0; 
        
    for Dimension=1:d 

      for NodeNumberTwo=1:NodeNumberMax
          
         for LineNumber=1:lineconfig(NodeNumberTwo,j) 
 
               if lineconfig(NodeNumberTwo,j)>0
           
                   NodeNumberMaxTwo=NodeNumberTwo; 
                   LineNumberTwo=LineNumber; 
                   
            internallinesval(LineNumber,Dimension,NodeNumberTwo,j)=(rand*kmax-kmax/2); 

   momentumsum(NodeNumberTwo,Dimension)=momentumsum(NodeNumberTwo,Dimension)+internallinesval(LineNumber,Dimension,NodeNumberTwo,j);
   
%                    test=internallinesval(LineNumber,Dimension,NodeNumber,j);

%             propagatorsval(LineNumber,Dimension,NodeNumber,j)=1/(test^2+m^2);

               end

           end       
     
       end    
        
     end
      
 end
     
     for NodeNumber=1:NodeNumberMax
     
         clear diff; diff(d)=0;  
           
           if NodeNumber>1  
           
               if NodeNumber<NodeNumberMax+1
               
             for Dimension=1:d 
               
              diff(Dimension)=momentumsum(NodeNumber-1,Dimension)-momentumsum(NodeNumber,Dimension);          
           
              momentumsum(NodeNumber,Dimension)=momentumsum(NodeNumber,Dimension)+diff(Dimension);  
              
             end
           
             for Dimension=1:d  
               
               internallinesval(LineNumberTwo,Dimension,NodeNumber,j)=diff(Dimension)+internallinesval(LineNumberTwo,Dimension,NodeNumber,j);  
               
             end
     
               end
             
           end 
           
           
           if NodeNumber==1  
               
               for s=1:ppoint  
                   
                 for Dimension=1:d 
                     
               internallinesval(LineNumberTwo,Dimension,NodeNumber,j)=internallinesval(LineNumberTwo,Dimension,NodeNumber,j)+kmomol(s,Dimension,samplex); 
           
               momentumsum(NodeNumber,Dimension)=momentumsum(NodeNumber,Dimension)+kmomol(s,Dimension,samplex);  
               
                 end 
                 
               end
               
           end 
           
 %          if NodeNumber==NodeNumberMax  
               
 %              for s=1:qpoint  
                   
 %                for Dimension=1:d 
                     
 %              internallinesval(LineNumberTwo,Dimension,NodeNumber,j)=internallinesval(NodeNumber,Dimension)+kmomor(s,Dimension,samplex); 
           
 %              momentumsum(NodeNumber,Dimension)=momentumsum(NodeNumber,Dimension)+kmomor(s,Dimension,samplex); 
               
 %                end 
                 
 %              end
               
              
 %            for Dimension=1:d 
               
 %             diff(Dimension)=-momentumsum(NodeNumber-1,Dimension)+momentumsum(NodeNumber,Dimension)-internallinesval(LineNumberTwo,Dimension,NodeNumber,j);          
           
 %             momentumsum(NodeNumber,Dimension)=momentumsum(NodeNumber,Dimension)+diff(Dimension)-internallinesval(LineNumberTwo,Dimension,NodeNumber,j);  
              
 %            end
           
 %            for Dimension=1:d  
               
 %             internallinesval(LineNumberTwo,Dimension,NodeNumber,j)=diff(Dimension);  
               
 %            end
     
 %         end 
          
          
%% end NodeNumberPerm          
          
end

%%  data of value of internal lines is stored in internallinesval  
%%  the value of the propagators is stored in the propagatorsval


%%  momentum conservation is chosen to change the last entry which isnt implemented yet 

%%   
%%  tree evaluation 
%%
%%    use a calling function   
%%

%%  at a node there are two sets of momenta 

%%   internallinesval(LineNumber,Dimension,NodeNumber,j)  
%%   internallinesval(LineNumber,Dimension,NodeNumber+1,j) 

%%   care must be taken to define the external momenta of the rainbow graphs 
%%     as momenta is defined from the left to the node 


     RainbowComputetwo=0; 

     
%% permutation set for the four point function 
     
   if permmomenta==1 
      
       kmomol;  kmomor; 

   end 

%% 1<->4

   if permmomenta==2  

       testvec(1,:,:)=kmomol(1,:,:); 
       testvec(2,:,:)=kmomor(2,:,:); 

       kmomol(1,:,:)=testvec(2,:,:); 
       kmomor(2,:,:)=testvec(1,:,:);  

   end 

%% 4<->3 which is an overall 1<->3 

   if permmomenta==3  

       testvec(1,:,:)=kmomol(1,:,:); 
       testvec(2,:,:)=kmomor(2,:,:); 

       kmomol(2,:,:)=testvec(1,:,:); 
       kmomor(1,:,:)=testvec(2,:,:);  

   end 


% gnumber=8; 

% ppoint=2; 

% qpoint=2; 

% select the left and right momenta from a set kmomol and kmomor

 clear kmomoleft; 
 clear kmomoright; 

kmomoleft(ppoint,d)=0; 
kmomoright(qpoint,d)=0;

%%%%% this is a subroutine to compute the rainbow iterations %%%%%%
%%
%% input external lines from the left side number p 
%%                               right side number q 
%%
%% with the number of coupling constants 
%%          this is the total number of internal lines 
%%
%% the output is the value of the graphs subject to the number of internal lines 
%%
%% summed over all their contributions 
%% 
%
%%%%% these are the inputs 

%% partition into all possible rainbow graphs 

%% npoint tree has gnumber=npoint-2  

%% rainbow graph with gnumber can have ppoint+2+qpoint+2+nodenumber*2 
%%                                        max number of couplings 
%%                          with all 4-point couplings in the internal nodes 

%%               with gnumber can have ppoint+qpoint+(gnumber-ppoint-qppoint-4) 
%%                          with only two nodes and no internal nodes  
%% 

% Two Nodes: InternalLines+ppoint-2+InternalLines+qpoint-2=gnumber  

InternalLinesMax=(gnumber+4-ppoint-qpoint)/2;  

InternalLinesMin=2;

% NodeNumber is nodes minus one

%%  this is the max number of nodes all internal line numbers are two 
%%  NodeNumberMax=(gnumber-ppoint-qpoint+2)/2 

NodeNumberMax=(gnumber-ppoint-qpoint+2)/2; 

RainbowTree=1.0; 

RainbowDenom=1.0; 

Atree=1.0; 

Denominator=1.0; 


for NodeNumber=1:NodeNumberMax-1  

       
RainbowTree=1.0;  

RainbowDenom=1.0; 

               leftno=lineconfig(NodeNumber,permno);  

               rightno=lineconfig(NodeNumber+1,permno);

          if rightno>0

          if leftno>0
              
             %% store in the number of lines to the right most node 


                    maxrightno=lineconfig(NodeNumber+1,permno);  

             %%

                        clear k;  k(leftno+rightno,d)=0; 
                        

            for LineNumber=1:leftno   

               for dindex=1:d 

         k(LineNumber,dindex)=internallinesval(LineNumber,dindex,NodeNumber,permno);
   
               end 

            end           

                     
            for LineNumber=1:lineconfig(NodeNumber+1,permno)  

                for dindex=1:d 

    k(LineNumber+leftno,dindex)=internallinesval(LineNumber,dindex,NodeNumber+1,permno);

                end 

            end 


   %%% this has setup the momenta variables in an array k_1 k_2 etc k_n 
   %%%   for the tree calling function  

   %%%   k_1 k_2 ... k_a  left    k_{a+1} ... k_n  right 


%% call in the numerical values of the trees  

       Atree=treecall(leftno+rightno,k,N,truenumbertree); 

       RainbowTree=RainbowTree*Atree; 

       Denominator=denomcall(leftno+rightno,k); 

       RainbowDenom=RainbowDenom*Denominator;   
       
       RainbowTree=RainbowTree*kmeasure^(leftno-1);

            end  
            
            end 
            
%% end the NodeNumber loop 

end 
  
 
%%  the two external trees have to be included 

%  first the left one   

          leftno=ppoint;  

          rightno=lineconfig(1,permno);  

          rightnodeno=1; 

          clear k; k(leftno+rightno,d)=0; 


         for i=1:leftno 

            for dindex=1:d 

                k(i,dindex)=kmomol(i,dindex,samplex); 

            end 

         end 


            for LineNumber=1:lineconfig(1,permno)   

               for dindex=1:d 

k(LineNumber+leftno,dindex)=internallinesval(LineNumber,dindex,rightnodeno,permno);
  
               end 

            end           

%% call in the numerical values of the left tree  

       Atree=treecall(leftno+rightno,k,N,truenumbertree); 

        RainbowTree=RainbowTree*Atree; 

        Denominator=denomcall(leftno+rightno,k); 

        RainbowDenom=RainbowDenom*Denominator; 
        
%  the right tree 


               rightno=qpoint; 

                      for i=1:NodeNumberMax 

                         if lineconfig(i,permno)>0; 

                             leftno=lineconfig(i,permno);  

                             leftnodeno=i; 

                          end 

                      end  


                        clear k;  k(leftno+rightno,d)=0; 


            for LineNumber=1:leftno

               for dindex=1:d 

   k(LineNumber,dindex)=internallinesval(LineNumber,dindex,leftnodeno,permno); 
               end 

            end           


         for i=1:qpoint 

            for dindex=1:d 

                k(i+leftno,dindex)=kmomor(i,dindex,samplex); 

            end 

         end


%% call in the numerical values of the right tree  


       Atree=treecall(leftno+rightno,k,N,truenumbertree); 

        RainbowTree=RainbowTree*Atree; 

        Denominator=denomcall(leftno+rightno,k); 

        RainbowDenom=RainbowDenom*Denominator; 

        RainbowTree=RainbowTree*kmeasure^(leftno-1);
        
%% this completes the tree nodal complex evaluation for the perm configuration

%% have to sum over the individual graphs in the nodal complex this is
%%    denoted by permno

% evaluate the graph 

% amputate the external lines

   denom=1; 

  for i=1:ppoint 

      denom=1/(kmomol(i,:,samplex)*kmomol(i,:,samplex)'+m^2)*denom; 

  end 

  for i=1:qpoint 

      denom=1/(kmomor(i,:,samplex)*kmomor(i,:,samplex)'+m^2)*denom; 
  
  end 


  RainbowIntegral(permno)=RainbowTree*RainbowDenom^(1/2)/denom^(1/2);  

% there is a z_2 sign ambiguity that needs to be determined from the 
% doubling of thepropagators in minkowski space

%% end of permutation set 

end 
  
%%  add up all the contributions from the multiple configurations 

RainbowIntegralScalar=0;  

for j=1:NodeNumberPerm 

   RainbowIntegralScalar=RainbowIntegralScalar+RainbowIntegral(j); 

end 

RainbowComputetwo=RainbowIntegralScalar+RainbowComputetwo; 


%% end of monte carlo simulation 

end 

%% end the permmomenta run 

end 

 RainbowCompute(samplex)=RainbowComputetwo;
 
%% end the samplex momenta run 

end

%%%%%%%%%%%%%%%%%%%%%%%%%%%%%%%%%%%%%%%%%%%
%%
%%  treecall.m
%%
%%

function[treecall]=treecall(n,k,N,truenumbertree)  

npoint=n; 

%%
%%  numerical values of trees 
%%%
%%    external momenta are inputs k(j,dindex) 
%% 
%%    npoint is the number of legs j=1,...,n 
% 
%%    N is the t_i^{[q]} matrix from which the trees are made  
%%

%%%%% defaults

m=1.0;

d=4;

TotTreeValue=0;  

clear kmominit;  kmominit(n,d)=0; 

clear kmomo;  kmomo(n,d)=0; 

clear kindex;  kindex(d)=0; 

pole=0;

%%%%%%%%%%%%%%%%%%%%%%%%%%%%%%%%%%%%%

for a=1:npoint

 for j=1:d

   kmominit(a,j)=k(a,j); 

   kmomo(a,j)=k(a,j); 

   kindex(j)=0;

 end

end

% input a Nmatrix and output a number 

% sum over the trees 

for j=1:truenumbertree(npoint) 

      kprod=1; 

   for a=1:npoint  

     for b=1:npoint-1 
        
for l=1:npoint

 for j=1:d

   kmominit(l,j)=k(l,j); 

   kmomo(l,j)=k(l,j); 

   kindex(j)=0;

 end

end
         
             pole=N(a,b,j,npoint); 
         
           if pole>0 

             kindex(:)=kmomo(a,:);

        for c=1:b 

           if a+c-1<npoint+1 

               for e=1:d
              
          kindex(e)=kmomo(a+c-1,e)+kindex(e);  

               end
         
           end 

           if a+c-1>npoint 

                for e=1:d
               
          kindex(e)=kmomo(a+c-1-npoint,e)+kindex(e); 

                end 
         
           end 

        end 

        kprod=kprod/(kindex*kindex'+m^2); 

           end

% storage of denominators as a function of steps 

     end 

   end 

   TreeValue=kprod;  

% add up the tree values 

   TotTreeValue=TotTreeValue+TreeValue; 

end

%% result for the treevalues is TotTreeValue 

%% rename to treecall 

treecall=TotTreeValue/2; 


%%%%%%%%%%%%%%%%%%%%%%%%%%%%%%%%%%%%%%%%%%%%%%%%%%%
%%
%%
%%   NodalComplex.m
%%
%%


function[NodalComplex,NodeNumberPerm]=NodalComplex(gnumber,ppoint,qpoint)


% clear data 
NodeNumber=1; 


NodeNumberMax=(gnumber-ppoint-qpoint+2)/2; 

clear lineconfig; clear lineconfigtwo; 

lineconfig(NodeNumberMax,1)=0; 

lineconfigtwo(NodeNumberMax,1)=0;

NodeNumberPerm=1;  

ltest=0; 

ptwo=0;

kSampleMax=400;


% Two Nodes: InternalLines+ppoint-2+InternalLines+qpoint-2=gnumber  

InternalLinesMax=(gnumber+4-ppoint-qpoint)/2;  

InternalLinesMin=2;

%% configs are stored in the array internallines(j) 
 
for NodeNumber=1:NodeNumberMax

 for ktwo=1:kSampleMax   
    
    clear internallines;

    for a=1:NodeNumber

       internallines(a)=0; 

    end 
    
    
%% a random sampling to determine the configs 

 for j=1:NodeNumber 

       InternalTest=0;

       while InternalTest<2; 

        InternalTest=round(rand*InternalLinesMax); 

          internallines(j)=InternalTest; 

        end 

 end 


addinternal=0; 

 for j=1:NodeNumber 

    addinternal=addinternal+internallines(j);  

 end 

%%   relation between gnumber and node line configuration 
%%
%%  ppoint+internalline1-2 + internalline1+internalline2-2 + ldots + 
%%        internalline(NodeNumber-1)+internallineNodeNumber-2 
%%          + internallineNodeNumber+qpoint-2 
%%
%% gnumber  = ppoint+qpoint-(NodeNumber+1)*2+\sum^{NodeNumber} 2*internallineJ
%% 

%% all double internal  gnumber=ppoint+qpoint-(NodeNumber+1)*2+NodeNumber*2*2 

%%                      NodeNumber=(gnumber-ppoint-qpoint+2)/2 

 addinternal=2*addinternal+ppoint+qpoint-2*(NodeNumber+1); 

    if addinternal==gnumber 
        
        ptwo=ptwo+1;
        
          for a=1:NodeNumber 
              
            lineconfigtwo(a,ptwo)=internallines(a);
         
          end
          
     end     
   
%% end random sample

  end

%% end NodeNumber loop

end 

% 190 to 312

%% eliminate the duplicates and store in lineconfig

 clear lineconfig; 

 lineconfig(NodeNumberMax,1)=0;  

 clear flagnumber; clear flagnumbertwo;
 
 flagnumber(ptwo)=0;  flagnumbertwo(ptwo)=0; 
 
   NodeNumberPerm=1; 
 
for r=1:ptwo-1 
    
   for s=r:ptwo
    
            testtwo=0; 
       
            for j=1:NodeNumberMax 
                
                testtwo=testtwo+abs(lineconfigtwo(j,r)-lineconfigtwo(j,s)); 
                
            end 
       
       if testtwo==0  
        
        %  if r<s 
            
              if flagnumbertwo(s)==0
              
              flagnumber(r)=1;  
              flagnumbertwo(s)=1; 

              end 
              
        % end     
              
      end
      
   end

end

 for r=1:ptwo 
     
     if flagnumber(r)==1 
         
         lineconfig(:,NodeNumberPerm)=lineconfigtwo(:,r)

         NodeNumberPerm=NodeNumberPerm+1; 
         
     end 

 end 

 NodeNumberPerm=NodeNumberPerm-1; 
 
%%% this completes the possible permutations of nodes and internal lines at
%%% this order   and   NodeNumberPerm is the number of permutations 

%%% all possible node configurations and internal lines are stored in 
%%     the variable 
%%% 
%%%                        lineconfig(NodeNumber,a)
%%%
%%%                       NodeNumber=node number
%%%
%%%                     a=configuration number up to NodeNumberPerm
%%
%%%


NodalComplex=lineconfig; 

NodalComplexPerm=NodeNumberPerm;


%%%%%%%%%%%%%%%%%%%%%%%%%%%%%%%%%%%%%%%%%%%%%%%%%%%
%%
%%
%%   NMatrix.m
%%
%%


function[NMatrix,truenumbertree]=NMatrix(npoint,npointlower)  

n=npoint; 

% npointlower is another useful delimiter

%% 
%%  t_i^{[q]} formation for the matrix of trees
%%
%%    npoint is the required dimension and is input  
%%
%%    matrix N is spanned by N(i,q,j,m) with m corresponding to m-point trees 
%%
%%       i,q represent t_i^{[q]} 
%%
%%       j corresponds to the different trees up to truenumbertree(m)
%%

clear N;

clear treenumber; 
treenumber(npoint)=0; 


for spoint=npointlower:npoint 

    rpoint=npoint-spoint+npointlower;  

    n=rpoint;


clear sigmanumber; clear flagnumber; 


  for k=1:Gamma(n+1)

          sigmanumber(1,k)=n;

    for j=2:n-2 

          sigmanumber(j,k)=0; 

          x=1;

        while x<2 

          x=round(n*rand);
     
          sigmanumber(j,k)=x;

        end 

    end 

  end 


% reorder the entries and eliminate redundancy


for k=1:Gamma(n+1)  

  for m=1:Gamma(n+1)

    for j=1:n-3 

      if sigmanumber(j,k)<sigmanumber(j+1,k) 

        test1=sigmanumber(j,k); test2=sigmanumber(j+1,k);  

        sigmanumber(j,k)=test2; sigmanumber(j+1,k)=test1; 
 
      end 

    end 

  end

end



%% the redundancy in sigmanumber is eliminated 

%% values are stored in rp  

%% note that the zero vector is allowed 

v=size(sigmanumber); dimp=v(2);

 for i=1:dimp 

  flagnumber(i)=0; 

 end



 for i=1:dimp 


% redundancy check  

   for m=1:dimp  


     if m==i 

        flagnumber(m)=1; 

     end



     if m~=i  

         if sigmanumber(:,m)==sigmanumber(:,i)  

           if i>m 

                flagnumber(m)=0;  

           else
                flagnumber(m)=1; 

           end 

         end 

      end 

   end 
  
 end


totalflagno=0;  

 for i=1:dimp 

     testmin=0;  

       for j=1:n-2 

        if sigmanumber(j,i)==2  

          testmin=testmin+1; 

        end 

       end 

       if testmin>1  

           flagnumber(i)=0;  

       end 

   totalflagno=totalflagno+flagnumber(i); 

 end 


rp(1,1)=0; clear rp


 for i=1:totalflagno

  for j=1:n-2

   rp(j,i)=0; 

  end 

 end

treenumbertwo=0;

k=0; 

 for i=1:dimp


% eliminate duplicate graphs 


   if flagnumber(i)==1  

      k=k+1; 

     rp(:,k)=sigmanumber(:,i);  

     treenumbertwo=treenumbertwo+1;   
  
   end 

 end

P=rp; 

treenumber(rpoint)=treenumbertwo; 

%%%%%%%%%%%%%%%%%%%%%%%%%%%%%%%
%%%%%%%%%%%%%%%%%%%%%%%%%%%%%%%
%%
%% 2) deduce the t_i^{[q]} variables 
%% 
%%    store in matrix T(i,q,treeno)
%% 

clear P;

clear M;

M(npoint,npoint,treenumber(npoint))=0; 


 for treeno=1:treenumber(rpoint)

  for i=1:n

    for q=1:n-1 

      M(i,q,treeno*rpoint)=0; 
    
    end 

  end 

 end 




for treeno=1:treenumber(rpoint)

%  for b=0:n-2

b=0;

%% first construct (p,[p_m]) sets, in P(i,j,treeno) matrix


  for j=1:n

      P(j,treeno)=0;  

      Qo(j,treeno)=0;

      PP(j,treeno)=0;

      Q(j,treeno)=0;

  end


  for j=1:n

    for i=1:n-2 

       if rp(i,treeno)==j 
 
            P(j,treeno)=P(j,treeno)+1;  
 
            Qo(j,treeno)=Qo(j,treeno)+1; 

            Q(j,treeno)=Qo(j,treeno);

       end 

       if j<n 

         if P(j,treeno)>0 

        for k=j+1:n 

          PP(k,treeno)=P(j,treeno); 

        end 

         end 

       end 

    end 

  end    

  for j=1:n 

    if P(j,treeno)==0 

       P(j,treeno)=PP(j,treeno); 

    end 

  end 



%% the numbers labeling the graphs are stored in rp(i,j) with first entry n
%% T is initialized to zero 


%% to find a pole n-1=number of nodes: search for subtrees   

%% scan all initial lines   

%%  finalline-initialine-1=nodes in tree between finalline and initialline


m=n;

nodes1=0; nodes2=0; 

b=0;

% permute the roundabout labeling as 1->permn  


      for j=1:n 

           if j+b<n+1 

         Q(j,treeno)=Qo(j+b,treeno);   

           end 
      
           if j+b>n 

         Q(j,treeno)=Qo(j+b-n,treeno); 

           end 

      end 


 for initialline=1:n-1 

    for finalline=initialline+1:n
 
                nodes1=0; nodes2=0;


                initcount=Q(initialline,treeno);

                finalcount=Q(finalline,treeno);  

 
                Qsum=0; 


% around clockwise the tree diagram

       if finalline-initialline>1

         if initcount==0 

         if finalcount>0

               for k=1:finalcount 

                   Qsum=0;

                 for alpha=initialline+1:finalline-1

                     Qsum=Qsum+Q(alpha,treeno);

                 end

                    Qsum=Qsum+k; 


          % skipping rule applied here  

               if Q(finalline-1,treeno)==0 
                   
          % check for subtree 
              

                     if Qsum==finalline-initialline

                        i=initialline; 

                        q=finalline-initialline+1;
  
                        M(i,q,treeno)=b+1; 

                     end 

                end 

                end

          end

          end

        end


    % finalline-initialline=1 case

             if finalline-initialline==1 

              if Q(initialline,treeno)==0 

                if Q(finalline,treeno)>0 

                   i=initialline; 

                   q=finalline-initialline+1; 

                        M(i,q,treeno)=b+1; 

                 end 

              end 

             end 


% end of initalline and finalline loop

  end 

 end

% end of the permutation

% end


% end the tree graph count 

end

% no end to the spoint loop yet 


% permute the matrices 

% wasteful of memory 

for i=1:rpoint

  treeno=treenumber(rpoint);

     for s=1:treenumber(rpoint)
           
    for a=1:rpoint 

      for b=1:rpoint 


         if a+i-1<rpoint+1 

             M(a+i-1,b,(i-1)*treeno+s)=M(a,b,s);  

         end 

         if a+i-1>rpoint 

             M(a+i-1-rpoint,b,(i-1)*treeno+s)=M(a,b,s); 
 
         end 

       end 

     end 

      end

end

% for treeno=1:rpoint*treenumber(rpoint)

   % for i=1:n 

   %    for q=1:n-2 

   %      if M(i,q,treeno)==1 

   %           if i+q<n 

   %        M(i+q+1,n-q,treeno)=0; 

   %           end 

   %           if i+q>n 

   %         M(i+q-n,n-q,treeno)=0; 

   %           end 

   %      end 

   %    end 

  %   M(i,rpoint,treeno)=0; M(i,1,treeno)=0; M(i,rpoint-1,treeno)=0; 

   % end 

   % end 


for j=1:rpoint*treenumber(rpoint) 

  for k=j+1:rpoint*treenumber(rpoint)



            test=0;

      for s=1:rpoint 

         for t=1:rpoint 

            test=abs(M(s,t,j)-M(s,t,k))+test; 

         end   

      end 

         if test==0

             if k>j 

           M(1,1,k)=M(1,1,k)+43*rand; 

             end 

         end 

        

  end 

end 


% there could be an error in the t_i^{[q]} matrix but not likely 


Nbefore(rpoint,rpoint,rpoint*treenumber(rpoint))=0;

N(rpoint,rpoint,rpoint*treenumber(rpoint),rpoint)=0;

nindex=0; 

for s=1:rpoint*treenumber(rpoint) 

     sumM=0; 

   for i=1:rpoint 

     for j=1:rpoint 

        sumM=sumM+M(i,j,s); 

     end 

   end 

      if sumM==rpoint-3  

           nindex=nindex+1; 

         Nbefore(:,:,nindex)=M(:,:,s); 

      end 

  
  N(:,:,:,rpoint)=Nbefore(:,:,:); 

  truenumbertree(rpoint)=nindex;  

end 


%%%  end loop on the rpoint index 

end 


%%%%%%%%%%%%%%%%%%%%%%%%%%%%%%%%%%%%%%%%%%
%%
%%  innerproduct.m
%%
%%


function[innerproduct]=innerproduct(ppoint,qpoint,kmomoleft,kmomoright)  

load Nmatrix 

   for j=1:ppoint
 
      test=test+kmomoleft(j,:)*kmomoright(j,:)'; 

   end

innerproduct=test; 


%%%%%%%%%%%%%%%%%%%%%%%%%%%%%%%%%%%%%%%%%  
%%
%%   denomcall.m
%%
%%


function[denomcall]=denomcall(numberlines,k) 


m=1.0; 

denom=1; 

d=4;

for i=1:d 

  p(i)=0; 

end 

   for j=1:numberlines 

      for i=1:d

         q(i)=k(j,i); 

      end

         denom=denom/(q*q'+m^2); 

    end 

denomcall=denom;


%%%%%%%%%%%%%%%%%%%%%%%%%%%%%%%%%%%%%%%%%%%%%%%%%%%%%%%%
%%%%%%%%%%%%%%%%%%%%%%%%%%%%%%%%%%%%%%%%%%%%%%%%%%%%%%%%


\end{lstlisting}

\end{document}